\documentclass[final,numberedheadings]{aipproc}

\usepackage{graphicx,amsmath,amsfonts}

\layoutstyle{8x11single}

\begin{document}

\title{Testing and selection cosmological models with dark energy}


\author{Marek Szyd{\l}owski}{
  address={Astronomical Observatory, Jagiellonian University, Orla 171, 30-244 Krak{\'o}w, Poland}
  ,altaddress={Mark Kac Complex Systems Research Centre, Jagiellonian University,\\ Reymonta 4, 30-059 Krak{\'o}w, Poland} 
}
\author{Aleksandra Kurek}{
  address={Astronomical Observatory, Jagiellonian University, Orla 171, 30-244 Krak{\'o}w, Poland}
}

\begin{abstract}
It is described dynamics of a large class of accelerating cosmological models in terms of dynamical systems of the Newtonian type. The evolution of the models is reduced to the motion of a particle in a potential well parameterized by the scale factor. This potential function can be reconstructed from distant supernovae type Ia data and many cosmological models represented in terms of the potential becomes in a good agreement with current observational data. It is proposed to use the information criteria to overcome this degeneracy within a class of A) dark energy models and B) simple models basing on modification of the FRW equation. Two class of models can be recommended by the Akaike (AIC) and Schwarz (BIC) information criteria: the phantom and $\Lambda$CDM models. 
\end{abstract}

\maketitle

\section{Introduction}

Recent measurements of distant supernovae (SNIa) \cite{Riess:1998cb,Perlmutter:1998np} indicate that our Universe is in an accelerating phase of expansion. This has stimulated the search for models explaining this phenomenon. There are two alternative types of explanation observed acceleration: 1) the Universe is accelerating due to presence of so called dark energy violating the strong energy condition \cite{Peebles:2002gy}, 2) instead of dark energy of unknown form the modification (or generalization) of the FRW equation is postulated. While the cosmological constant is still attractive proposition for dark energy, it offers only some effective model without understanding of observed value of vacuum energy. This situation open the cosmology on different propositions of candidates for dark energy description \cite{Padmanabhan:2002ji,Copeland:2006wr}. All these models can be described in terms of a two-dimensional dynamical system of the Newtonian type if it is assumed a  general form of the equation of state $p_{X}=w(a)\rho_{X}$ for dark energy $X$ \cite{Szydlowski:2003cf,Szydlowski:2003fg,Szydlowski:2005bn}. From the conservation condition (under condition spatial homogeneity and isotropy)
\begin{equation}\label{eq:1}
 \dot{\rho}=-3H(\rho + p),
\end{equation}
where $\rho$ and $p$ are effective energy density and pressure respectively, $H=(\ln a)^{\cdot}$ is the Hubble function, $a$ is the scale factor, a dot denotes the differentiation with respect to the cosmological time. We obtain that $\rho=\rho(a)$ or $\rho=\rho(z)$, where $(1+z)=a^{-1}$ ($a_{0}$- the present value of the scale factor is chosen as a unity).

From the Raychaundhuri equation (called also the acceleration equation) we obtain additional equation governed evolution of the Universe
\begin{equation}\label{eq:2}
\frac{\ddot{a}}{a}= - \frac{1}{6}(\rho + 3p).
\end{equation}
One can simply check that equation (\ref{eq:2}) assumes the form analogous to the form of the Newtonian equation of motion $\ddot{a}=-\frac{\partial V}{\partial a}$ if we choose the potential in the form
\begin{equation}\label{eq:3}
V = - \frac{\rho a^{2}}{6},
\end{equation}
where $\rho$ plays the role of effective energy density which satisfies the conservation condition.

Equation (\ref{eq:2}) admits first integral
\begin{equation}\label{eq:4}
\frac{(\dot{a})^{2}}{2} + V(a) = E,
\end{equation}
which is called the Friedmann first integral. Finally, due to the particle-like description of the evolution of the Universe, which mimics motion of the fictitious particle moving in the potential well. The potential function is representing single function of scale factor which determine full evolutional scenarios of the cosmological model. Therefore our proposition is using scale factor $a$ and its derivative as independent variables which measure states of the system. Moreover in such approach we have additionally Hamiltonian structure and the FRW cosmological system is representing in terms of two-dimensional dynamical system. In such formulation space of states is defined in the following way:
\begin{equation}\label{eq:5}
\zeta = \left\{ (a, \dot{a} ) \colon \frac{(\dot{a})^{2}}{2} + V(a) = -\frac{k}{2} \right\},
\end{equation}
where $k$ is the curvature index, $k=0,\pm 1 $. $\zeta$ should be treated as a phase space of the system in which trajectories lies on algebraic curves which are determined by the Hamiltonian constrained $\mathcal{H}= \frac{(\dot{a})^{2}}{2} + V(a) = -\frac{k}{2}$. If we reparameterize time variable $t \mapsto \tau \colon |H_{0}|dt = d\tau \ (H_{0}\neq 0)$ then $\zeta$ can be defined equivalently as:
\begin{equation}\label{eq:6}
\zeta = \left\{ (x, x') \colon \frac{(x')^{2}}{2} + V(x) = -\frac{\Omega_{k,0}}{2}\right\},
\end{equation}
where $ \Omega_{k,0}= \frac {-3 k}{(a_{0})^{2}}/3(H_{0})^{2}$ is the density parameter for curvature fluid, $x=\frac{a}{a_{0}}$ is a dimensionless scale factor expressed in its present value $a_{0}$. If we assume that fluid which filled the Universe is composed with noninteracting fluids: dust matter, radiation and the cosmological constant then we obtain:
\begin{equation}\label{eq:7}
\frac{(x')^{2}}{2} + V(x) = \frac{\Omega_{k,0}}{2} 
\end{equation}
where
\[
V(x)=-\frac{1}{2}\{ \Omega_{\text{m},0} x^{-1} + \Omega_{\Lambda,0} x^{2}+ \Omega _{r,0}x^{-2} \}.
\]
Of course all density parameters $\Omega_{i,0}$ are not independent and satisfy the constraint condition $\sum_{i=(m,\ r,\ k)} \Omega_{i,0} =1$.

The original Friedmann first integral (\ref{eq:4}) can be rewritten to the form
\begin{equation}\label{eq:8}
\left( \frac {dx}{d\tau} \right) ^{2}= \Omega_{\text{m},0}(x^{-1}-1) + \Omega_{r,0}(x^{-2}-1)+ \Omega_{\Lambda,0} ( x^{2}-1) + 1.
\end{equation}
In order to survey the set of all its solutions (which have an arbitrary parameter) we pick a pair $(\Omega_{\text{m},0}, \Omega_{r,0})$ and represent the range of each possible solution $x(\tau)$ corresponding to these data and an initial value $\Omega_{\Lambda,0}$, as a horizontal segment in a plane $(x,\ \Omega_{\Lambda,0})$. For this purpose it is necessary to mark the boundary of the permitted region  $(x>0,\ x^{2} \ge 0 \ \text{and}\ V(x)\le \frac{1}{2}\Omega_{k,0})$. The boundary is corresponding to $V=\frac{1}{2}\Omega_{k,0}$ and $a=0$ ($\Omega_{\Lambda,0}$ - axis). From $V(x)= \frac{1}{2}\Omega_{k,0},\\ (x'^{2}=0)$ we obtain:
\begin{equation}\label{eq:9}
\Omega_{\Lambda,0}(x) = \frac{x(x-1)\Omega_{\text{m},0} + (x-1)(x+1)\Omega_{r,0}-x^{2}}{x^{2}(x+1)(x-1)}.
\end{equation}
From this boundary curve it is possible to obtain qualitatively a classification of all evolutional scenario in the configuration space. Each model corresponding to a maximally extended line segment $\Omega _{\Lambda,0}=\textrm{const}$ which includes the present value $x=1$. If that segment ($\Omega _{\Lambda,0}(x)=\textrm{const}$ level) is bounded on the left by the locus $a=0$, the model has a big bang. The potential function is given by
\begin{equation}\label{eq:10}
V(x)=-\frac{1}{2}\frac {x^{2}(x^{2}-1) \Omega_{\Lambda,0} + x(1-x)\Omega_{\text{m},0} + (1-x^{2})\Omega_{r,0} + x^{2}}{x^{2}},
\end{equation}
and all trajectories lies in the classically allowed region $V(x)\le \frac{1}{2}\Omega_{k,0}$.

Let us consider now a more general model filled by dark energy satisfying the equation of state $p_{X}=w_{X}(a)\rho_{X}$ instead of the cosmological constant $\Lambda$. Then we obtain the potential function in the form 
\begin{equation}\label{eq:11}
V(x)=-\frac{1}{2}\frac {x^{2}(f(x)-1) \Omega_{X,0} + x(1-x)\Omega_{\text{m},0} + (1-x^{2})\Omega_{r,0} + x^{2}}{x^{2}}
\end{equation}
where
\[
\Omega_{X}=\Omega_{X,0} f(x),\ f(x)=x^{-1} \exp\left[-3\int_{1}^{x}\frac{w_{X}(a)}{a}da\right].
\]
The full classification of dark energy models can be performed in the analogous way like to lambda models from consideration of a boundary curve $V(x)= \frac{1}{2}\Omega_{k,0}$.

\section{Ensemble of dark energy models}

Table 1 shows some examples of different dark energy models appeared in cosmological investigations sources of acceleration. In Table 1 we present also potentials function for alternative propositions of explanation acceleration without conception of dark energy. For these class of models it is proposed some modification of the Friedmann first integral (so called the Cardassian models) or new brane physics is incorporated (Dvali-Gabadadze-Porrati (DGB) models or the Randall-Sundrum brane cosmological models). 

Let us introduce the definition of basic notion. By the ensemble of the FRW cosmological models with dark energy it is understood a subspace of all dynamical systems on the plane which are represented the FRW dynamical systems of a Newtonian type, characterized by the potential function $V(a)$. This space can be naturally equipment in structure of the Banach space if we introduce the $C^{1}$ metric $d(1,2)=\max_{x \in E} \{|V_{1}-V_{2}|,\ |V_{1x}-V_{2x}| \}$, where $E$ is a closed subspace of the configuration space, $V_{1}$ and
$V_{2}$ are potential functions labeled as $1\& 2$.
\begin{table}
\begin{tabular}{lcll}
\hline
  & \tablehead{1}{c}{b}{case}
  & \tablehead{1}{c}{b}{name of model}
  & \tablehead{1}{c}{b}{form of potential function}\\
\hline
 A & 1 & models with matter and & $ V(x) = -\frac{1}{2} \left( \Omega_{\text{m},0}x^{-1}+ \sum_{i} \Omega_{i,0}x^{-1-3w_{i}}\right)$ \\
   &   & noninteracting fluids  & \\
   &   & $p_{i}=w_{i}\rho_{i},\ i=1,\dots,n $ & \\
   &   &                                      &\\
   & 2 & phantom models & $ V(x) = -\frac{1}{2} \left( \Omega_{\text{m},0}x^{-1}+\Omega_{X,0}x^{3}\right) $ \\
   &   & $p_{X}=-\frac{4}{3}\rho_{X}$ and dust & \\
   &   &  &\\ 
   & 3 & models with generalized & $V(x) = -\frac{1}{2} \left[\Omega_{\text{m},0}x^{-1} + \Omega_{\text{Ch},0}x^{2}\left( A_{s}+\frac{1-A_{s}}{x^{3(1+\alpha)}}\right)^{\frac{1}{1+\alpha}}\right] $ \\
   &    & Chaplygin gas &\\
   &    &   &\\ 
   & 4 & models with dynamical E.Q.S. & $V(x) = -\frac{1}{2} \left\{ \Omega_{\text{m},0}x^{-1}+\Omega_{X,0}x^{-1-3(w_{0}+w_{1})}\exp\left[ 3w_{1}\left( \frac{1}{x} -1 \right) \right] \right\} $\\
   &    &$p_{X}=(w_{0}+w_{1}z)\rho_{X}$& \\ 
\hline
 B & 5 & RSB models & $V(x) = -\frac{1}{2} \left( \Omega_{\text{m},0}x^{-1}+\Omega_{dr,0}x^{-2}+\Omega_{\lambda,0}x^{-4}+\Omega_{\Lambda,0}x^2 \right)$ \\
   &   & &\\
   & 6 & Cardassian models & $V(x) = -\frac{1}{2} \left( \Omega_{\text{m},0}x^{-1}+\Omega_{\text{C},0}x^{-3n+2} \right)$\\
   &   & $\rho=\rho_{\text{m},0}x^{-3}+\rho_{C,0}x^{-3n}$ &\\
   &   & &\\
   & 7 & DGP brane models & $V(x) = -\frac{1}{2} \left(\sqrt{\Omega_{\text{m},0}x^{-1}+\Omega_{rc,0}x^{2}}+ \sqrt{\Omega_{rc,0}x^{4}}\right)$ \\
   &   & &\\
   & 8 & bouncing models & $V(x) = -\frac{1}{2} \left( \Omega_{\text{m},0}x^{-m+2}-\Omega_{n,0}x^{-n+2}+\Omega_{\Lambda,0}x^{2} \right)$\\
\hline
\end{tabular}
\caption{Different models explaining acceleration of the Universe in terms of dark energy (A) and modification of the FRW equation (B). All models are defined in terms of a single potential function.}
\end{table}

The Bayesian approach to the hypothesis testing was developed by Jeffreys \cite{Jeffreys:1935,Jeffreys:1961} as a program of methodology for quantifying the evidence of two competing scientific theories. In our context we test hypotheses about origins of the present acceleration of the Universe. In Jeffreys' approach the crucial role plays statistical models which are used to represent the probability of the data. The Bayes theorem is used to calculate the posterior probabilities that one of the models is correct or to find the most favored model. 

The characteristic feature of Bayesian framework is that it was applied in several fields of sciences. The Bayesian methodology offers the possibility of comparison of nonnested models and accounts for uncertainty in the choice of models \cite{Kass:1995}.

\section{Idea of model selection}
In the development of cosmology the basic role played an idea of cosmological models together with an idea of astronomical tests \cite{Ellis:1999sx}. The idea of cosmological tests make cosmological models parts of astronomy which offers possibility of observationally determining the set of realistic parameters, that can characterize viable models. While we can perform estimation of model parameters using a standard minimization procedure based on the likelihood method, many different scenarios are still in a good agreement with observational data of SNIa. This problem which appears in observational cosmology is called the degeneracy problem. To solve this problem it is required to differentiate between different dark energy models. We propose to use the Akaike information criterion (AIC) and Bayesian information criterion (BIC).

For the model selection framework it is required to have data and a set of models and then the model based statistical inference. The model selection should be based on a well-justified single (even naive) model or, at least, a simple model which suffices for making inferences from the data. In our case the $\Lambda$CDM model plays just the role of such a model and application of information criteria seems to be reasonable as a first step toward  an extension of the standard maximum likelihood method. The model selection should be viewed as a way to obtain model weights, not just a way to select only one model (and then ignore that selection occurred). Moreover in the notion of true models do not believe information theories because the model by definition is only approximations to unknown physical reality: There is no true model of the Universe that perfectly reflect large structure of space-time but some single best model has been found.

In this paper the Bayesian factor is used to select among dark energy models. The Bayesian factor is the posterior odds of the one hypothesis when the prior probabilities of two alternative hypotheses are equal. It can be defined as the ratio of the marginal likelihoods under two competing models, i.e. the probabilities of the data with all the model parameters integrated out. Thus the marginal likelihood is a basic quantity (evidence) needed for this approach \cite{Burnham:2002}. We use the Bayesian factor for evidence provided by SNIa data in favor of one cosmological model with dark energy (represented by statistical model) as opposed to another. The well known the Bayesian information criterion (BIC) gives the approximation to the logarithm of the Bayes factor and can be easily calculated without the requirement of prior distribution. It is necessary to apply this method when a very large number of different dark energy models is considered and we wish to select only one of them. This problem is solved in the Bayesian framework by choosing the model with the highest posterior probability. 

The AIC is defined in the following way \cite{Akaike:1974}
\begin{equation}\label{eq:12}
\text{AIC}=-2\ln\mathcal{L} + 2d,
\end{equation}
where $\mathcal{L}$ is the maximum likelihood and $d$ is a number of model parameters. The best model with a parameter set providing the preferred fit to the data is that which minimizes the AIC. While there are justification of the AIC in information theory and also rigorous statistical foundation for the AIC, it can be also justified as Bayesian using a `savvy' prior on models that is a function of a sample size and a number of model parameters. For the AIC we can define $\Delta$AIC$_{i}$ as the difference of the AIC for model $i$ and the AIC value for the scale model: $\Delta_{i}=\text{AIC}_{i} - \text{AIC}_{1}$. The $\Delta_{i}$ are easy to interpret and allow a quick `strength of evidence' comparison and a ranking of candidates for dark energy description. The models with $\Delta_{i}\le 2$ have substantial support (evidence), those where $2<\Delta_{i}\le 7$ have considerably less support, while models having $\Delta_{i} > 10 $ have essentially no support.

In the Bayesian framework a best model (from the model set $\{M_{i}\}$) is that which has the largest value of probability in the light of data (so called posteriori probability): $P(M_{i}|D)=\frac{P(D|M_{i})P(M_{i})}{P(D)}$, where $P(M_{i})$ is a priori probability for the model $M_{i}$, $P(D|M_{i})$ is the evidence ($E_{i}$). We can define the posterior odds for models $M_{i}$ and $M_{j}$, which (in the case when no model is favored a priori) is reduced to the evidence ratio (so called the Bayes factor - $B_{ij}$).

Schwarz \cite{Schwarz:1978} showed that for observations coming from linear exponential family distribution: $\ln E_{i}=\ln\mathcal{L} - \frac{1}{2}d \ln N +O(1)$, where $N$ is number of data points used in the fit (this was extended by Haughton \cite{Haughton:1988} for curved exponential family). According to this result Schwarz introduced a criterion for the model selection: the best model is that which minimized the BIC, defined as
\begin{equation}\label{eq:13}
\text{BIC}=-2 \ln \mathcal{L} + d \ln N.
\end{equation}
To compare models $M_{i}$ and $M_{j}$ one can compute $2 \ln B_{ij} = - (\text{BIC}_{i} - \text{BIC}_{j}) \equiv - \Delta \text{BIC} _{ij}$ which can be interpret as `strength of evidence' against $j$ model: $0\leq 2\ln B_{ij} < 2$--not worth more than a bare mention, $2\leq 2\ln B_{ij} < 6$ -- positive, $6\leq 2\ln B_{ij} < 10$ -- strong, and $2\ln B_{ij}\ge 10$ -- very strong.

There are many simulation studies in the statistical literature on either the AIC or BIC alone, or often comparing them and making recommendation on which one to use. It should be pointed out that assumptions of using classical model selection methods are satisfied, namely:
\begin{enumerate}
\item there is model-based inference from SNIa data (luminosity distance observable)
\item there is a set of dark energy models and no certainty about which model should be used in explanation of present acceleration
\item a data-based choice is made among these competing models (see Table 1)
\end{enumerate}
Then the inference is made from this one selected model as if it were a priori the only model fit to the data.

Table 2 gives the value of the AIC and BIC for flat models from Table 1.
It also contains values of $\Delta\textrm{AIC}_{i1}$ and $2\ln B_{1i}$, where model $1$ is our scale model ($\Lambda$CDM).
\begin{table}
\begin{tabular}{ccccr@{,}lr@{,}l}
\hline
  & \tablehead{1}{c}{b}{Case}
  & \tablehead{1}{c}{b}{AIC ($\Omega_{k,0}=0)$}
  & \tablehead{1}{c}{b}{BIC ($\Omega_{k,0}=0)$}
  & \tablehead{2}{c}{b}{$\Delta\textrm{AIC}_{i1}$} 
  & \tablehead{2}{c}{b}{$2 \ln B _{1i}$}\\  
\hline
 A&1 ($w_{X}=-1$)      & 179,9 & 186,0 & 0&0&0&0    \\
  &2                & 178,0 & 184,1 & -1&9 & -1&9 \\
 &3                & 181,7 & 193,9 & 1&8 & 7&9 \\
 &4                & 180,5 & 192,7 & 0&6 & 6&7 \\
\hline
B&5&182,1&194,3&2&2&8&3\\
 &6&178,5&187,7&-1&4&1&7\\
 &7&180,9&187,0&1&0&1&0\\
 &8&183,9&196,2&4&0&10&2\\              
\hline
\end{tabular}
\caption{Values of the AIC, BIC, $\Delta\textrm{AIC}_{i1}$ and $2\ln B_{1i}$ for flat models from Table 1.}
\end{table}
 
\begin{figure}
  \includegraphics[height=.26\textheight]{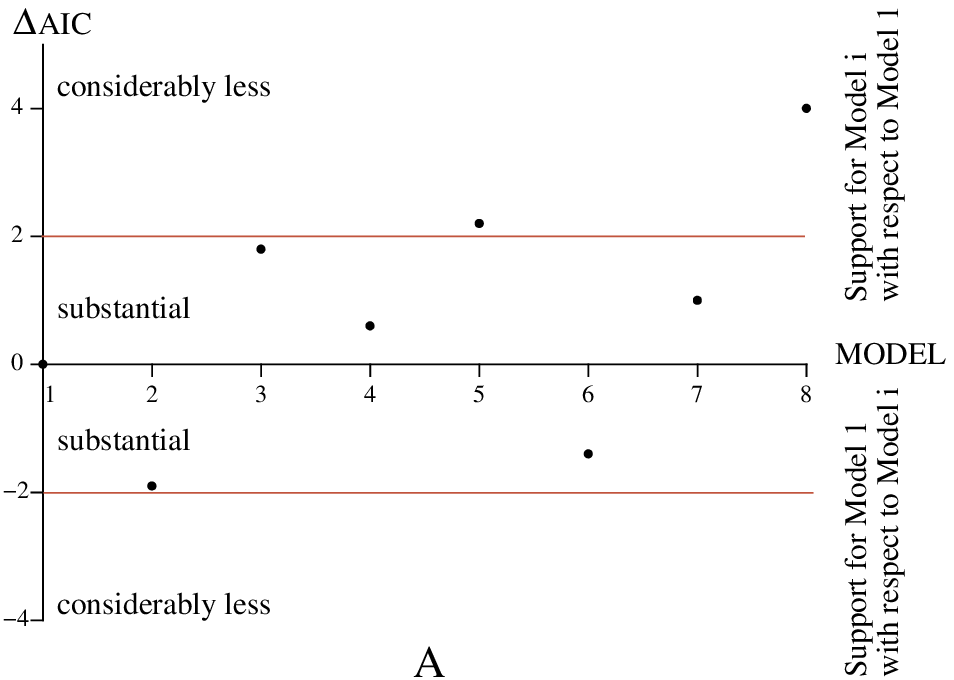}
  \includegraphics[height=.26\textheight]{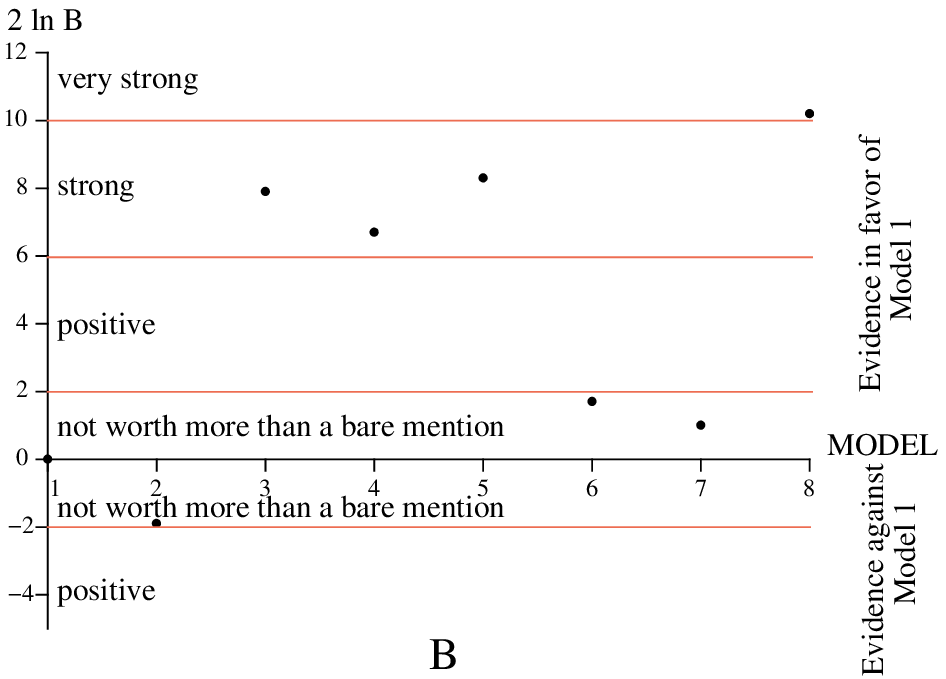}
  \caption{Value of A) $\Delta\textrm{AIC}_{i1}= \textrm{AIC}_{i}-\textrm{AIC}_{1}$ and B) $2\ln B_{1i}$  for models from Table 1.}
\end{figure}

From results contained in Table 2 one can conclude that the phantom model (case $2$) is the best model from our model set, but the $\Lambda$CDM model has still substantial support (evidence). The Akaike information criterion indicates that there is only one strong disfavored model (case $8$), whereas the Bayesian information criterion denotes that also cases $3,4$ and $5$ are less probable than the $\Lambda$CDM model (see Figure 1).

\begin{theacknowledgments}
The paper was supported by Astro-Lea PF Polish-French 2005 program. The authors are very grateful to participants of seminar on observational cosmology, especially to Dr W. Godlowski for helpful discussion. M. Szydlowski is also grateful to Prof. Roland Triay for fruitful discussion.     
\end{theacknowledgments}


\end{document}